\documentclass[12pt,a4paper]{article}
\usepackage{amsmath}
\usepackage{latexsym}
\usepackage{amssymb}
\usepackage{graphicx,color}
\makeatletter
\def\rddots{\mathinner{\mkern1mu\raise\p@%
    \vbox{\kern7\p@\hbox{.}}\mkern2mu%
    \raise4\p@\hbox{.}\mkern2mu\raise7\p@\hbox{.}\mkern1mu}}
\makeatother
\makeatletter
\def\eqnarray{%
\stepcounter{equation}%
\let\@currentlabel=\theequation
\global\@eqnswtrue
\global\@eqcnt\z@
\tabskip\@centering
\let\\=\@eqncr
$$\halign to \displaywidth\bgroup\@eqnsel\hskip\@centering
$\displaystyle\tabskip\z@{##}$&\global\@eqcnt\@ne
\hfil$\displaystyle{{}##{}}$\hfil
&\global\@eqcnt\tw@$\displaystyle\tabskip\z@{##}$\hfil
\tabskip\@centering&\llap{##}\tabskip\z@\cr}
\makeatother
\newcommand{\fukuso}{{\mathbf C}}

\begin{document}

\title{\sl An Approximate Solution of the Master Equation with 
the Dissipator being a Set of Projectors}
\author{
  Kazuyuki FUJII
  \thanks{E-mail address : fujii@yokohama-cu.ac.jp }\\
  Department of Mathematical Sciences\\
  Yokohama City University\\
  Yokohama, 236--0027\\
  Japan
  }
\date{}
\maketitle
%
%
%
%
\begin{abstract}
  In this paper we consider a quantum open system and treat the master 
  equation with some restricted dissipator which consists of a set of 
  projection operators (projectors). The exact solution is given under 
  the commutable approximation (in our terminology). This is the first 
  step for constructing a general solution. 
\end{abstract}
%


%
%
%
%

\vspace{10mm}
In this paper we revisit dynamics of a quantum open system. First of all 
we explain our purpose in a short manner. See \cite{BP} as a general 
introduction to this subject. 
We consider a quantum open system $S$ coupled to the environment $E$. 
Then the total system $S+E$ is described by the Hamiltonian
\[
H_{S+E}=H_{S}\otimes {\bf 1}_{E}+{\bf 1}_{S}\otimes H_{E}+H_{I}
\]
where $H_{S}$, $H_{E}$ are respectively the Hamiltonians of the system and 
environment, and $H_{I}$ is the Hamiltonian of the interaction.

\par \noindent
Then under several assumptions (see \cite{BP}) the reduced dynamics of the 
system (which is not unitary !) is given by the Master Equation
\begin{equation}
\label{eq:master-equation}
\frac{\partial}{\partial t}\rho=-i[H_{S},\rho]-{\cal D}(\rho)
\end{equation}
with the dissipator being the usual Lindblad form
\begin{equation}
\label{eq:dissipator}
{\cal D}(\rho)=\frac{1}{2}\sum_{\{j\}}
\left(A_{j}^{\dagger}A_{j}\rho+\rho A_{j}^{\dagger}A_{j}
-2A_{j}\rho A_{j}^{\dagger}\right).
\end{equation}
Here $\rho\equiv \rho(t)$ is the density operator (matrix) of the system. 

It is not easy to solve the equation (\ref{eq:master-equation}) with 
the dissipator (\ref{eq:dissipator}), so we make a simple and 
convenient assumption. Namely, the generators $\{A_{j}\}$ are given by 
$A_{j}=\sqrt{\lambda_{j}}P_{j}$ with projectors $\{P_{j}\}$ ; 
$P_{j}^{\dagger}=P_{j},\ P_{j}^{2}=P_{j},\ P_{j}P_{k}=\delta_{jk}P_{k}$. 
Note that we don't assume the rank $P_{j}$ =1 (extended models). 
Then the dissipator becomes
\begin{equation}
\label{eq:simple-dissipator}
{\cal D}(\rho)=\frac{1}{2}\sum_{\{j\}}\lambda_{j}
\left(P_{j}\rho+\rho P_{j}-2P_{j}\rho P_{j}\right)
\end{equation}
where $\{\lambda_{j}\}$ are decoherence parameters to determine the 
strength of the interaction. See \cite{BG}, \cite{KD} (in \cite{KD} there is 
a very compact description on this subject). 
It is interesting to rewrite (\ref{eq:simple-dissipator}) as
\begin{equation}
\label{eq:simple-dissipator-2}
{\cal D}(\rho)=\frac{1}{2}\sum_{\{j\}}\lambda_{j}
\left\{P_{j}\rho ({\bf 1}-P_{j})+({\bf 1}-P_{j})\rho P_{j}\right\}
\equiv 
\frac{1}{2}\sum_{\{j\}}\lambda_{j}
\left(P_{j}\rho Q_{j}+Q_{j}\rho P_{j}\right).
\end{equation}
Note that $\{Q_{j}\}$ are also projectors satisfying $P_{j}Q_{j}=
Q_{j}P_{j}=0$ for $j\in \{j\}$.

As a result we have only to solve the equation
\begin{equation}
\label{eq:master-equation-2}
\frac{\partial}{\partial t}\rho=-i(H\rho-\rho H)-
\frac{1}{2}\sum_{\{j\}}\lambda_{j}
\left(P_{j}\rho Q_{j}+Q_{j}\rho P_{j}\right)
\end{equation}
where we have set $H=H_{S}$ for simplicity.

In order to attack the equation (\ref{eq:master-equation-2}) let us 
make some mathematical preliminaries. For a matrix $X=(x_{ij})\in 
M(n;\fukuso)$ we correspond to the vector 
$\widehat{X}\in {\fukuso}^{n^{2}}$ as
\begin{equation}
\label{eq:correspondence}
X=(x_{ij})\ \longrightarrow\ 
\widehat{X}=(x_{11},x_{12},\cdots,x_{1n},\cdots \cdots,x_{n1},x_{n2},
\cdots,x_{nn})^{T}
\end{equation}
where $T$ means the transpose. Then the following formula is well--known
\begin{equation}
\label{eq:well--known formula}
\widehat{AXB}=(A\otimes B^{T})\widehat{X}
\end{equation}
for $A,B,X\in M(n;\fukuso)$. Since the proof is easy we leave it to 
readers. 

By use of the formula the equation (\ref{eq:master-equation-2}) 
can be rewritten as
\begin{equation}
\label{eq:master-equation-3}
\frac{\partial}{\partial t}\widehat{\rho}=
\left\{
-i(H\otimes {\bf 1}-{\bf 1}\otimes H^{T})-
\frac{1}{2}\sum_{\{j\}}\lambda_{j}
\left(P_{j}\otimes Q_{j}^{T}+Q_{j}\otimes P_{j}^{T}\right)
\right\}
\widehat{\rho},
\end{equation}
therefore the formal solution is given by
\begin{equation}
\label{eq:formal-solution}
\widehat{\rho}(t)=
\exp
\left\{
-it(H\otimes {\bf 1}-{\bf 1}\otimes H^{T})-
t\sum_{\{j\}}(\lambda_{j}/2)
\left(P_{j}\otimes Q_{j}^{T}+Q_{j}\otimes P_{j}^{T}\right)
\right\}
\widehat{\rho}(0).
\end{equation}

To calculate $\exp(\cdots)$ explicitly is (almost) impossible, so 
we must appeal to some approximation method. For that let us remind 
the Baker--Campbell--Hausdorff (B-C-H) formula. 
For $A,B\in M(n;\fukuso)$ we want to decompose as
\begin{equation}
\label{eq:decomposition}
\mbox{e}^{A+B}=\mbox{e}^{A}\mbox{e}^{I(A,B)}\mbox{e}^{B}.
\end{equation}
The ``interaction" term $I(A,B)$ is given by
\begin{equation}
\label{eq:interaction term}
I(A,B)=-\frac{1}{2}[A,B]+\frac{1}{6}\left\{[[A,B],B]+[A,[A,B]]\right\}
+\cdots .
\end{equation}
The proof is easy. In fact,  
$
\mbox{e}^{I(A,B)}=\mbox{e}^{-A}\mbox{e}^{A+B}\mbox{e}^{-B}
$
 by (\ref{eq:decomposition}) and we have only to apply 
the B-C-H formula (\cite{Vara} and see also \cite{FS} as an 
interesting topic)
\[
\mbox{e}^{X}\mbox{e}^{Y}=
\mbox{e}^{X+Y+(1/2)[X,Y]+(1/12)\{[[X,Y],Y]+[X,[X,Y]]\}+\cdots}
\quad 
\mbox{for}
\
X,Y \in M(n;\fukuso)
\]
two times.

For 
\[
A=-it(H\otimes {\bf 1}-{\bf 1}\otimes H^{T}),\quad 
B=-t\sum_{\{j\}}(\lambda_{j}/2)
\left(P_{j}\otimes Q_{j}^{T}+Q_{j}\otimes P_{j}^{T}\right)
\]
there is no method to calculate $\mbox{e}^{I(A,B)}$ explicitly 
as far as we know. Therefore we ignore this term, namely let us 
call it the ``commutable approximation".

Under the commutable approximation we have only to calculate

\begin{eqnarray}
\label{eq:approximate-solution}
\widehat{\rho}(t)
&\approx&
\exp
\left\{-it(H\otimes {\bf 1}-{\bf 1}\otimes H^{T})\right\}
\exp
\left\{
-t\sum_{\{j\}}(\lambda_{j}/2)
\left(P_{j}\otimes Q_{j}^{T}+Q_{j}\otimes P_{j}^{T}\right)
\right\}
\widehat{\rho}(0) \nonumber \\
&=&
\left(\mbox{e}^{-itH}\otimes \mbox{e}^{itH^{T}}\right)
\exp
\left\{
-t\sum_{\{j\}}(\lambda_{j}/2)
\left(P_{j}\otimes Q_{j}^{T}+Q_{j}\otimes P_{j}^{T}\right)
\right\}
\widehat{\rho}(0) \nonumber \\
&=&
\left(\mbox{e}^{-itH}\otimes \left(\mbox{e}^{itH}\right)^{T}\right)
\exp
\left\{
-t\sum_{\{j\}}(\lambda_{j}/2)
\left(P_{j}\otimes Q_{j}^{T}+Q_{j}\otimes P_{j}^{T}\right)
\right\}
\widehat{\rho}(0).
\end{eqnarray}

Next let us calculate the second term in (\ref{eq:approximate-solution}), 
which is not so difficult as follows.
\begin{eqnarray}
\label{eq:step-1}
(\sharp)
&\equiv&
\exp
\left\{
-t\sum_{\{j\}}(\lambda_{j}/2)
\left(P_{j}\otimes Q_{j}^{T}+Q_{j}\otimes P_{j}^{T}\right)
\right\} \nonumber \\
&=&
\prod_{\{j\}}
\exp
\left\{
(-\lambda_{j}t/2)
\left(P_{j}\otimes Q_{j}^{T}+Q_{j}\otimes P_{j}^{T}\right)
\right\} \nonumber \\
&=&
\prod_{\{j\}}
\left\{
{\bf 1}\otimes {\bf 1}+\left(\mbox{e}^{-\lambda_{j}t/2}-1\right)
\left(P_{j}\otimes Q_{j}^{T}+Q_{j}\otimes P_{j}^{T}\right)
\right\}
\end{eqnarray}
where we have used facts

\par \noindent
(a)\ \ $\left\{P_{j}\otimes Q_{j}^{T}+Q_{j}\otimes P_{j}^{T}\ |\ 
j\in {\{j\}}\right\}$ are projectors commuting with each other.

\par \noindent
(b)\ \ $\mbox{e}^{\lambda R}={\bf 1}+
\left(\mbox{e}^{\lambda}-1\right)R$\ \ if $R$ is a projector.

Here we set $R_{j}=P_{j}\otimes Q_{j}^{T}+Q_{j}\otimes P_{j}^{T}$. 
For $i < j < k$ we obtain

\par \noindent
(c)\ \ $R_{i}R_{j}=\left(P_{i}\otimes Q_{i}^{T}+Q_{i}\otimes P_{i}^{T}
\right)\left(P_{j}\otimes Q_{j}^{T}+Q_{j}\otimes P_{j}^{T}\right)=
P_{i}\otimes P_{j}^{T}+P_{j}\otimes P_{i}^{T}$.

\par \noindent
(d)\ \ $R_{i}R_{j}R_{k}=
\left(P_{i}\otimes P_{j}^{T}+P_{j}\otimes P_{i}^{T}\right)
\left(P_{k}\otimes Q_{k}^{T}+Q_{k}\otimes P_{k}^{T}\right)
=0$.

\par \noindent
From (\ref{eq:step-1}) and (c), (d)
\begin{eqnarray}
\label{eq:step-2}
(\sharp)
&=&
\prod_{\{j\}}
\left\{
{\bf 1}\otimes {\bf 1}+\left(\mbox{e}^{-\lambda_{j}t/2}-1\right)
R_{j}
\right\} \nonumber \\
&=&
{\bf 1}\otimes {\bf 1}+
\sum_{j}\left(\mbox{e}^{-\lambda_{j}t/2}-1\right)R_{j}+
\sum_{j<k}
\left(\mbox{e}^{-\lambda_{j}t/2}-1\right)
\left(\mbox{e}^{-\lambda_{k}t/2}-1\right)R_{j}R_{k}
\nonumber \\
&=&
{\bf 1}\otimes {\bf 1}+
\sum_{j}\left(\mbox{e}^{-\lambda_{j}t/2}-1\right)
\left(P_{j}\otimes Q_{j}^{T}+Q_{j}\otimes P_{j}^{T}\right)+
\nonumber \\
&{}&
\sum_{j<k}
\left(\mbox{e}^{-\lambda_{j}t/2}-1\right)
\left(\mbox{e}^{-\lambda_{k}t/2}-1\right)
\left(P_{j}\otimes P_{k}^{T}+P_{k}\otimes P_{j}^{T}\right).
\end{eqnarray}

Therefore
\begin{eqnarray}
\label{eq:step-3}
\widehat{\rho}(t)
\approx && 
\left(\mbox{e}^{-itH}\otimes \left(\mbox{e}^{itH}\right)^{T}\right)
\left\{
{\bf 1}\otimes {\bf 1}+
\sum_{j}\left(\mbox{e}^{-\lambda_{j}t/2}-1\right)
\left(P_{j}\otimes Q_{j}^{T}+Q_{j}\otimes P_{j}^{T}\right)+
\right.
\nonumber \\
&{}&
\left.
\sum_{j<k}
\left(\mbox{e}^{-\lambda_{j}t/2}-1\right)
\left(\mbox{e}^{-\lambda_{k}t/2}-1\right)
\left(P_{j}\otimes P_{k}^{T}+P_{k}\otimes P_{j}^{T}\right)
\right\}\widehat{\rho}(0).
\end{eqnarray}
Coming back to matrix form by use of (\ref{eq:well--known formula}) 
we finally obtain
\begin{eqnarray}
\label{eq:final-1}
{\rho}(t)\approx 
\mbox{e}^{-itH}
\left\{
{\rho}(0)+\sum_{j}\left(\mbox{e}^{-\lambda_{j}t/2}-1\right)
\left(P_{j}{\rho}(0)Q_{j}+Q_{j}{\rho}(0)P_{j}\right)+
\right. \nonumber \\
\left.
\sum_{j<k}
\left(\mbox{e}^{-\lambda_{j}t/2}-1\right)
\left(\mbox{e}^{-\lambda_{k}t/2}-1\right)
\left(P_{j}{\rho}(0)P_{k}+P_{k}{\rho}(0)P_{j}\right)
\right\}
\mbox{e}^{itH}
\end{eqnarray}
or
\begin{eqnarray}
\label{eq:final-2}
{\rho}(t)\approx 
\mbox{e}^{-itH}
\left\{
{\rho}(0)+\sum_{j}\left(\mbox{e}^{-\lambda_{j}t/2}-1\right)
\left(P_{j}{\rho}(0)Q_{j}+Q_{j}{\rho}(0)P_{j}\right)+
\right. \nonumber \\
\left.
\frac{1}{2}
\sum_{j\ne k}
\left(\mbox{e}^{-\lambda_{j}t/2}-1\right)
\left(\mbox{e}^{-\lambda_{k}t/2}-1\right)
\left(P_{j}{\rho}(0)P_{k}+P_{k}{\rho}(0)P_{j}\right)
\right\}
\mbox{e}^{itH}
\end{eqnarray}
for $j,k\in \{j\}$. This is the main result.

\vspace{3mm}
A comment is in order.\ In the two qubit system a general density matrix 
is written as
\[
\rho(t)=\frac{1}{4}
\left(
{\bf 1}_{2}\otimes {\bf 1}_{2}+
p_{i}(t){\sigma}_{i}\otimes {\bf 1}_{2}+
q_{j}(t){\bf 1}_{2}\otimes {\sigma}_{j}+
r_{ij}(t)\sigma_{i}\otimes \sigma_{j}
\right)
\]
where we have used the Einstein's notation on summation . Using this 
expression one tries to solve the equation coming from pure 
decoherence term
\[
\frac{\partial}{\partial t}\rho
=
-\frac{1}{2}\sum_{\{j\}}\lambda_{j}
\left(P_{j}\rho+\rho P_{j}-2P_{j}\rho P_{j}\right)
=
-\frac{1}{2}\sum_{\{j\}}\lambda_{j}
\left(P_{j}\rho Q_{j}+Q_{j}\rho P_{j}\right).
\]
The equation is then reduced to a set of (relatively simple) 
equations of $\{p\}$, $\{q\}$ and $\{r\}$. 
However, such a method (trial) is irrelevant as shown in the paper. 
Our method is quite general !

\vspace{5mm}
In this paper we considered the master equation with the dissipative 
being a set of projectors and constructed the exact solution under 
the commutable approximation. This is just the first step for 
constructing a general solution for the equation.

In order to take one step forward we must take the ``interaction" 
term $I(A,B)$ in (\ref{eq:interaction term}) into consideration. 
However, such a method to calculate it has not been known as far as 
we know. Therefore it may be reasonable to restrict our target to 
some simple models. 
Further work will be needed and we will report it in a near future, 
\cite{KF1}.

On the other hand we are studying some related topics, see \cite{SR} 
and \cite{KF2}. However, we make no comment on them in the paper.

Lastly, we conclude the paper by stating our motivation. We are studying 
a quantum computation (computer) based on Cavity QED (see \cite{FHKW1} 
and \cite{FHKW2}), so to construct a more realistic model 
of (robust) quantum computer we have to study a severe problem coming 
from decoherence. 
This is our future task.

\vspace{5mm}
\noindent{\em Acknowledgment.}\\
The author wishes to thank K. Funahashi for helpful comments and suggestions.


\end{document}